\documentclass[aapm, mph,
 amsmath,amssymb,
 reprint,
]{revtex4-1}

\usepackage{graphicx}% Include figure files
\usepackage{dcolumn}% Align table columns on decimal point
\usepackage{bm}% bold math

\usepackage[english]{babel}
\usepackage[utf8]{inputenc}
\usepackage{epsfig,amssymb,latexsym,amsmath,pifont}
\newcommand{\p}{\varphi}
\newcommand{\e}{\epsilon}

\newcommand{\om}{\omega}
\usepackage{graphicx}
\usepackage{soul}
\usepackage{braket}
\usepackage{mathtools}

\begin{document}

%\preprint{}

\title{Synchronization in time-varying random networks with vanishing connectivity} %Title of paper



\author{Marco Faggian}
\affiliation{SUPA, Physics Department and Institute for Complex Systems and Mathematical Biology, King's College, University of Aberdeen, AB24 3UE (UK)} 
\affiliation{Faculty of Information Studies in Novo Mesto, 8000 Novo Mesto (Slovenia)}

\author{Francesco Ginelli}
\affiliation{SUPA, Physics Department and Institute for Complex Systems and Mathematical Biology, King's College, University of Aberdeen, AB24 3UE (UK)} 

\author{Fernando Rosas}
\affiliation{Centre of Complexity Science and Department of Mathematics, Imperial College London (UK)} 
\affiliation{Department of Electrical and Electronic Engineering, Imperial College London (UK)}

\author{Zoran Levnaji\'c}
\affiliation{Faculty of Information Studies in Novo Mesto, 8000 Novo Mesto (Slovenia)}


\date{\today}

\begin{abstract}
A sufficiently connected topology linking the constituent units of a complex system is usually seen as a prerequisite for the emergence of collective phenomena such as synchronization. We present a random network of heterogeneous phase oscillators in which the links mediating the interactions are constantly rearranged with a characteristic timescale and, possibly, an extremely low instantaneous connectivity. We show that, provided strong coupling and fast enough rewiring are considered, the network is able to reach partial synchronization even in the vanishing connectivity limit.
We also provide an intuitive analytical argument, based on the comparison between the different characteristic timescales of our system in the low connectivity regime, which is able to predict the transition to synchronization threshold with satisfactory precision. In the formal fast switching limit, finally, we argue that the onset of collective synchronization is captured by the time-averaged connectivity network.
Our results may be relevant to qualitatively describe the emergence of consensus in social communities with time-varying interactions and to study the onset of collective behavior in engineered systems of mobile units with limited wireless capabilities.
\end{abstract}


\maketitle %\maketitle must follow title, authors, abstract and \pacs


\section{Introduction}

The emergence of collective phenomena in complex systems is related to the interplay between interaction topology and local dynamics \cite{VespignaniBook,newman,costa, barabasi}. Stationary connections can lead to coherent dynamical patterns, typically studied in the framework of network theory, with the local dynamics taking places on individual nodes and interactions modelled as links.
In the context of complex networks, conditions of minimal connectivity are know for enabling the emergence of collective dynamics \cite{VespignaniBook, mason}. A prominent example is {\it synchronization} in networks of oscillators \cite{arkadybook,Arenas2008,arkadyandme,arkadymisha}, where the connectivity thresholds for a wide range of different network topologies have been determined in great detail \cite{Rodrigues2016,acebron}. However, many complex systems, and particularly social and engineered ones, may not maintain a constant connectivity, but rather yield a topology defined by a time-dependent connectivity matrix $\mathcal{A}_{ij}^t$. Examples range from animal groups \cite{Ermentrout91, Ballerini08} and time-dependent plasticity in neural networks \cite{Markram97, Maistrenko07} to robot swarms \cite{Pini2011}, human social networks \cite{Sekara16} and communication networks of moving units \cite{Hua2009}.

Synchronization in time-varying networks received considerable attention in the control and nonlinear dynamics literature \cite{Hasler04, Stilwell06, Amritkar06, Li08, Lucas18}. Yet, these efforts almost exclusively concentrated on systems composed of homogeneous units, largely relying on standard linear stability analysis. Introducing quenched disorder, that is, considering systems composed by many heterogeneous oscillators is however more challenging, especially for a finite number of units, as the stability of the {\it partially synchronized state} cannot be typically treated by simple linear stability analysis \cite{Strogatz2000}. In this situation, averaging theorems \cite{AveragingT} may not be trivially applicable, so that different approaches may be needed.

A first step towards the study of time-varying networks of heterogeneous oscillators is provided by Ref. \cite{So08}, which focused on two populations of oscillators switching between two fixed topologies at a given frequency. Interestingly, analysis of this ``blinking'' network revealed that high-frequency switching may induce synchronization, even when the two individual topologies can only sustain an incoherent phase. 

While these findings provide a first hint that results for time-varying networks of homogeneous units can be extended to the heterogeneous case, here we wish to take a step further and study a time-varying network of heterogeneous oscillators, where individual nodes interact randomly (and possibly quite seldomly) in both time and oscillator space
In particular, here we ask under which conditions macroscopic synchronization may emerge in Erd\"os-R\'eny networks with random rewiring and arbitrarily small instantaneous connectivity.  We thus consider $N$ heterogeneous agents interacting randomly, with a bidirectional and typically {\it sparse} connectivity matrix in a regime of strong coupling. We will see that our system is characterized by three different timescales: interacting agents quickly converge towards a common state on a short {\it local syncronization} timescale $\tau_{LS}$, while each agent may randomly rewire all his connections with a typical {\it rewiring} timescale $T$. When two connected agents are separated, their internal states diverge, with yet another {\it local de-sinchronization} timescale $\tau_{LD}$ which depends on their heterogeneity and it is typically larger than $\tau_{LS}$. 
One can interpret this setup as a crude model of social interactions, where individuals interact in time with different subsets of their common social network. When interacting, and despite their intrinsic differences, they tend to quickly converge towards a common opinion, but when separated, their differences take over again and their opinion diverges. 

Model parameters allow to control the separation between these characteristic times, which enables a detailed study of the emergence of synchronization in relation to the interplay between different timescales. In the following, we show that -- provided the links are rewired frequently enough -- dynamics can permanently achieve a partially synchronized state, even when the instantaneous connectivity is far smaller than what is needed to synchronize stationary networks. In particular, via numerical simulations and approximate analytical arguments of a concrete model, we show that for a sufficiently strong coupling and a sufficiently fast rewiring, our system reaches and maintains a macroscopic (partially) synchronized state, even in the limit of vanishing connectivity: it is the high frequency blinking of links that prevents the system from relaxing into an incoherent state as would happen with stationary topologies. 

This paper is organized as follows: in Section \ref{sec2} we define a precise model for our time-varying network and sketch its synchronization phase diagram through direct numerical simulations. In Section \ref{sec3} we focus on the low connectivity regime. Analysing the characteristic timescales of the system and invoking an averaging theorem in the limit $T\to0$, we provide an approximate expression for the synchronization threshold which compares favorably with numerical estimates. In Section \ref{sec4} we first discuss higher connectivities, where the instantaneous network topology is characterized by a system spanning giant component, and then argue that the transition to synchronization belongs to the standard Kuramoto class in the entire phase diagram. Conclusions are finally drawn in Section \ref{sec5}.

\section{Kuramoto model on time-varying networks}
\label{sec2}
\subsection{Model definition}

We first introduce our model. 
Let us consider a network of $N$ Kuramoto oscillators, where the state of the $i$-th node is represented by a phase variable $\varphi_i \in [0,2\pi]$. Each oscillator is characterized by a quenched natural frequency $\omega_i$, drawn from a zero-mean Gaussian distribution with standard deviation $\sigma$. 
Oscillators interact with each others according to a time-varying adjacency matrix $\mathcal{A}_{ij}^t$, with $m_i^t=\sum_j \mathcal{A}_{ij}^t$ being the instantaneous degree of node $i$. For simplicity, we chose the adjacency matrix to be symmetric and with binary values $\mathcal{A}_{ij}^t=0,1$, leaving other cases for future studies. Hence, the dynamics of the oscillators obey the following equation:
\begin{equation}
\dot{\varphi}_i = \omega_i + \frac{\varepsilon}{m_i^t} \sum_{j} \mathcal{A}_{ij}^t(T)\sin(\varphi_j - \varphi_i) \; ,
\label{maineq}
\end{equation}
where epsilon quantifies the strength of the coupling \cite{NOTE1}. Obviously, when no edges at all insist on node $i$ we have simply $\dot{\p}_i = \om_i$.

The dynamics of $\mathcal{A}_{ij}^t(T)$ is determined as follows. At each moment, the adjacency matrix corresponds to a random, or Erd\"os-R\'eny (ER) network, defined by the vertex number $N$ and the linking probability $p$ \cite{VespignaniBook}. The random rewiring of edges is then modelled as a Poissonian process, with each individual node rewiring synchronously all its incident edges with probability rate $1/T$, with $T$ being the typical {\it rewiring time}. In the instantaneous rewiring process of vertex $i$, all edges incident on $i$ are first deleted; all the potential links of vertex $i$ are then considered, and new edges $i-j$ are created with probability $p$. 
It is well known that the topological properties of ER networks are essentially determined by the mean degree connectivity $\langle m \rangle = (N-1) p$, so that in the following we find convenient to define $q=p\,N \approx \langle m \rangle $ and adopt $q \approx \langle m \rangle$ as the relevant connectivity parameter.

Note finally that Eq. (\ref{maineq}) is invariant under the following rescaling: 
\begin{equation}
\begin{cases}
t'=\alpha t\\
\sigma'=\frac{\sigma}{\alpha}\\
\varepsilon'=\frac{\varepsilon}{\alpha}\\
\end{cases}
\label{scaling}
\end{equation}
(with  $\alpha\in\mathbb{R}^+$), provided also the rewiring time is rescaled accordingly, $T'=\alpha T$.
Due to this invariance, it is easy to show that the dynamics is actually controlled by the two dimensionless quantities $T/\sigma$ and $\varepsilon/\sigma$ and by the connectivity $q$.

One can interpret this setup as a crude model for several natural/social phenomena. Consider for example social interactions, where individuals interact in time with different subsets of their common social network, according to a certain frequency of personal encounters/interactions. When interacting, despite intrinsic differences of their opinions (i.e. different quenched natural frequencies), individuals tend to quickly converge towards a common opinion. However, when separated, their differences take over again, and their opinions slowly diverge. The spectrum of natural frequencies $\omega_i$ can thus represent the range of ``unperturbed'' opinions of a population, while in Eq. (\ref{maineq}) interactions with other persons (nodes) leads to the effective frequencies $\omega^{eff}_i (t) = \dot{\varphi}_i$, representing the actual opinion of the agents.

It is well known that the Kuramoto model with stationary network connectivity, either globally connected \cite{Strogatz2000} or with other sufficiently connected topologies \cite{Rodrigues2016}, displays a synchronized solution for large enough couplings $\e$. In this synchronized state, a macroscopic fraction of oscillators share a common effective frequency, reaching macroscopic consensus in our point of view.
The degree of synchronization can be evaluated through the standard instantaneous Kuramoto order parameter
\begin{equation}
R(t) = \left| \frac{1}{N} \sum_{k=1}^N e^{i\p_k (t)} \right| \; , 
\end{equation}
which is finite for synchronized states and tends to zero as $1/\sqrt{N}$ in the absence of macroscopic synchronization. 
In the following, we will typically consider its average over time and disorder (i.e. different natural frequencies realization), $\Delta = \braket{R}_{t\,,\omega}$, and make use of its different finite size scaling behavior to better estimate the transition between (partial) synchronization and disorder.

\subsection{Direct numerical simulations}
\label{sims}

\begin{figure}[t!]
\centering
\includegraphics[width=0.9\linewidth]{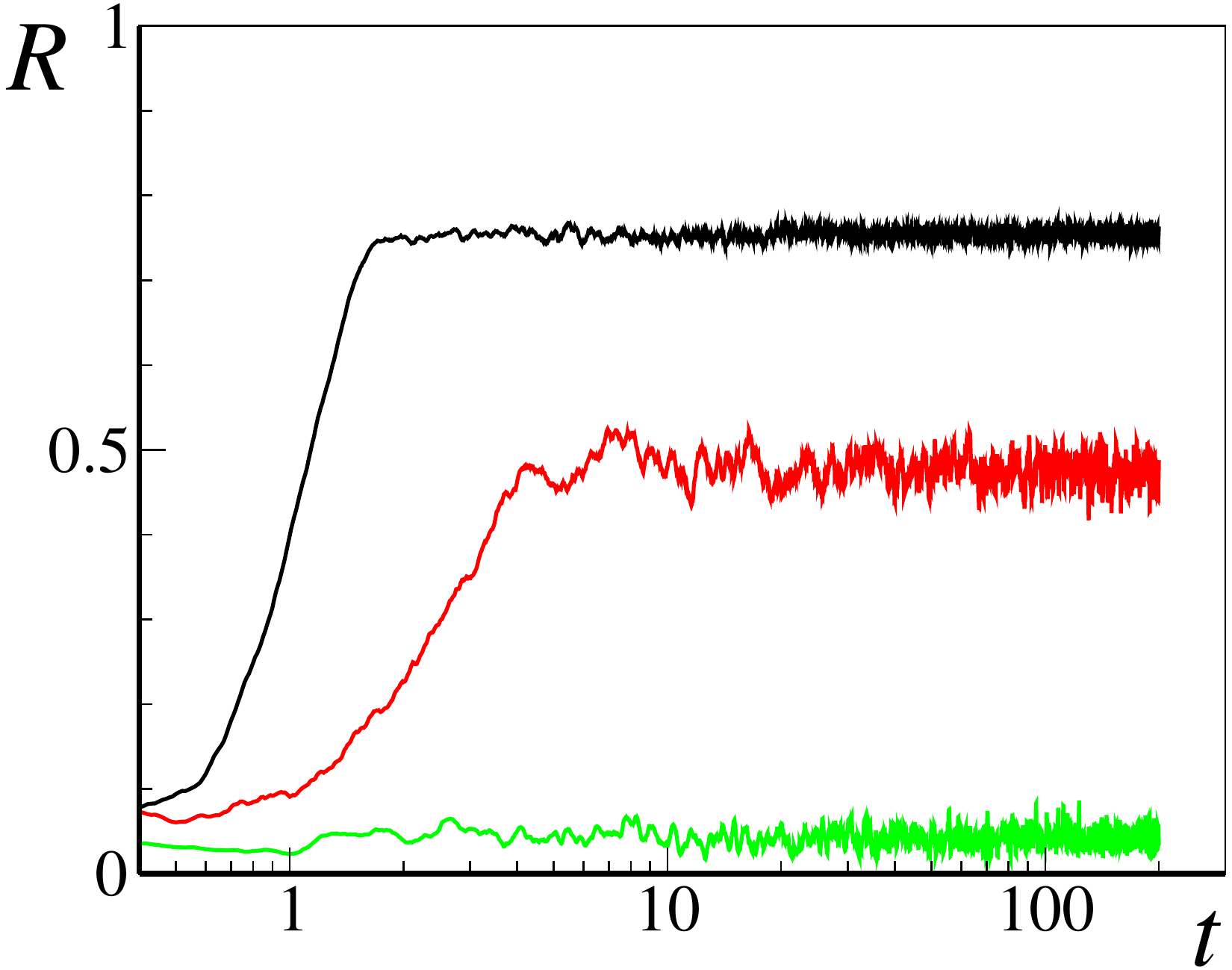}
%\vspace{0.5cm}
\caption{Lin-log plot of the order parameter $R$ as function of time for a network with $N=10^4$ and $q=0.8$, with $\varepsilon=8$ and $\sigma=1$. The three curves correspond to three different values of switching time $T$: green $T=6.28$, red $T=0.63$, and black $T=0.31$.}
\label{figure1}
\end{figure}

\begin{figure}[t!]
\centering
\includegraphics[width=0.99\linewidth]{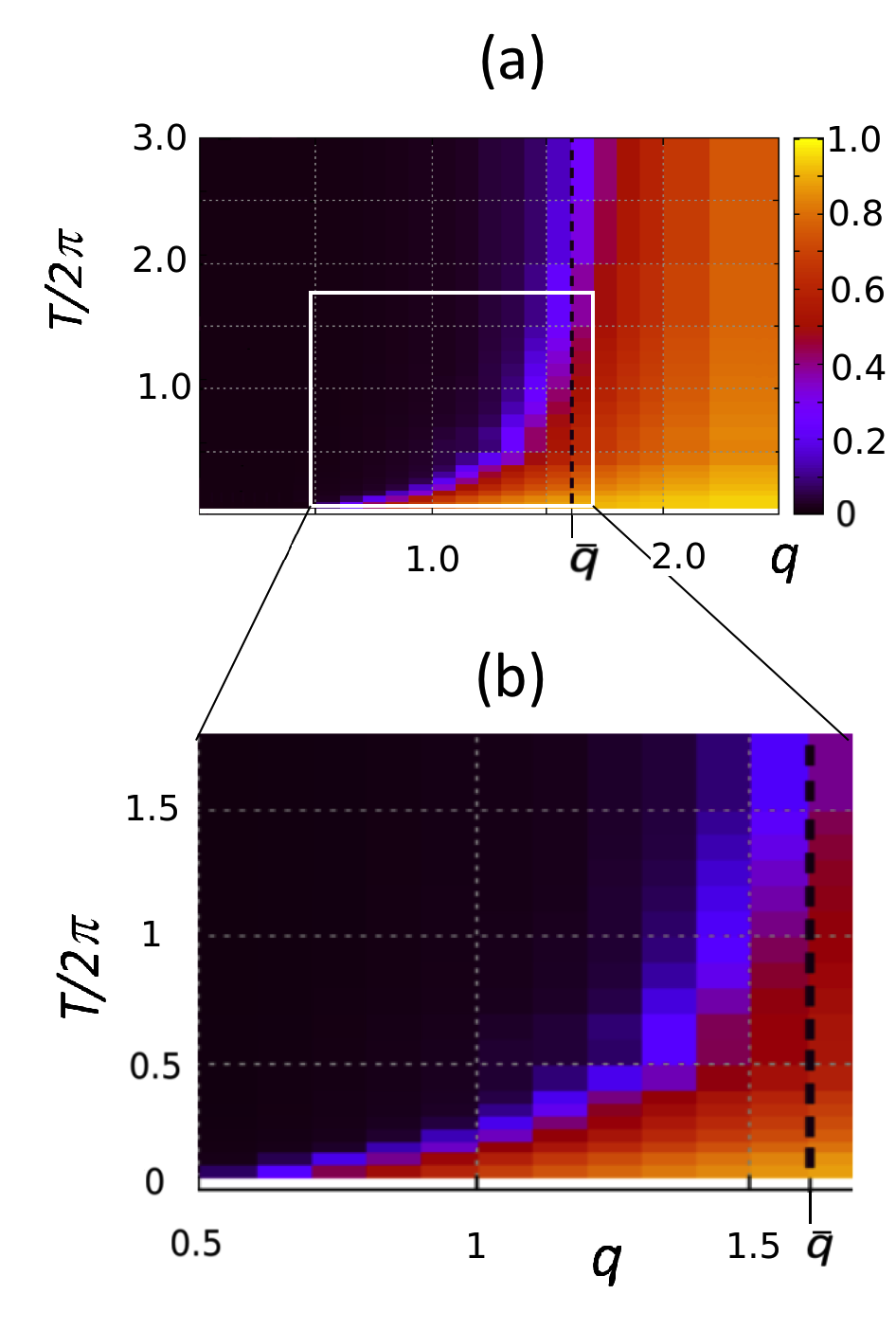}
\caption{ (a) Stationary values of the order parameter $\Delta$ (color-coded according to the right vertical bar) as function of the rewiring frequency ($T$) and network connectivity ($q$). Simulations have been performed for a network of $N=10^4$ Kuramoto oscillators with $\varepsilon=8$ and $\sigma=1$. Values have been averaged over $\Omega=10$ different realizations. For small values of $q$ the system is strongly dependent on the value of the rewiring time $T$ and the phase diagram shows a clear transition from partial synchronization to disorder as $T$ is increased beyond a critical value $T_c(q)$. At larger $q$ values, the transition approaches a vertical asymptote, roughly located at $q=\bar{q}=1.66(6)$ (dashed black line). For $q>\bar{q}$ the dynamics achieves partial synchronisation regardless of the value of $T$.
(b) Zoomed view of panel (a) in the range $q\in [0.5, 1.7]$.}
\label{figure2} 
\end{figure}

In this work, numerical simulation are performed using a standard 4th order Runge-Kutta integrator of step $dt$. After each Runge-Kutta time-step, each vertex may undergo a rewiring event (as defined above) with Poissonian probability 
\begin{equation}
r=1-\exp(-dt/T) \;.
\end{equation}
We use a time-step of at most $dt=10^{-2}$. When investigating fast network dynamics however, we are forced to adopt time-steps smaller than the network rearrangement timescale $T$, 
that is $dt \approx T/10$. 

In order to illustrate the behavior of our time-varying network dynamics in a strong coupling regime, $\e=8$, $\sigma=1$, we begin presenting numerical simulations of the dynamics (\ref{maineq})  for a network of $N=10^4$ elements and a mean connectivity $q=0.8$. 

As it is shown in Fig.\ref{figure1}, no synchronization emerges when the rewiring is sufficiently slow ($T \approx 6.3$ in this example). As the rewiring time is lowered past a synchronization threshold, we observe macroscopic synchronization with an increasing order parameter $R(t)$. 
This shows that sufficiently fast network rewiring can overcome the effects of low network connectivity, inducing partial synchronization on the network.
 
We next want to characterize with more details the parameter space $(q,T)$. We do that by repeating the above computation for a grid (lattice) of different values of $q$ and $T$. For each of these values we calculate $\Delta = \braket{R}_{t\,,\omega}$ by averaging over 10 random realizations of the quenched natural frequencies $\omega_i$ and different random initial phases. Time averages are performed over the stationary part of $R(t)$, after a proper initial transient has been discarded. The results are shown via colorplot in Fig.\ref{figure2}a (the lighter the color, the larger the value of $\Delta$). We first observe that for a sufficiently large connectivity, $q>\bar{q}$, the system always reaches macroscopic synchronization, regardless of the rewiring time $T$.
Analysis of the averaged order parameter $\Delta$ in the large $T$ limit, as reported in more details in Section \ref{largeC}, suggests $\bar{q} = 1.66(6)$.
Here, we just wish to point out that $\bar{q}$ is clearly larger than $q=1$, the threshold for the emergence of a giant connected component in ER graphs \cite{giantcomponent}. 
In this regime, synchronization is indeed to be expected in the strong coupling limit, due to sufficient interactions among the oscillators.  

For smaller values of $q$, on the other hand, where no large components characterize the instantaneous network topologies, sufficiently fast rewiring is needed to achieve synchronization, at least for $q>0.5$. A transition line $T_c(q)$ separating partial synchrony from incoherence (i.e. the violet border between the dark and the bright zone) in the plane $(q, T)$ can be roughly identified from this colour plot. Indeed, a closer look at the phase diagram, as reported in Fig.\ref{figure2}b, suggests that the transition line $T_c(q)$ separating partial synchrony from incoherence in the plane $(q, T)$ is initially characterized by a linear behaviour. For larger connectivity values, on the other hand, $T_c(q)$ grows faster than linear, finally diverging as a vertical asymptote is approached at $q=\bar{q}$. Note however that the transition line is characterized by a non zero intercept at $q=q_0\approx 0.5$ with the $T=0$ axis. Thus, for smaller connectivity values ($q \lessapprox 0.5$), no synchronization is possible for the coupling $\e/\sigma = 8$, no matter how fast is the rewiring. 

In the following section we will proceed to better characterize the transition to synchrony in the low average connectivity region $q<1$ by means of approximate analytical arguments and detailed numerical simulations.

%---------------------------------------------------------------------------------

\section{Synchronization for low and vanishing connectivity}
\label{sec3}
\subsection{Characteristic time scales and the onset of synchronization}
\label{scales}

We next seek to understand the physical mechanism leading to synchronisation in the low connectivity region via switching. For this we need to grasp the three characteristic timescales governing information flow and the dynamics of our system. The first time scale is the \textit{local synchronisation} time $\tau_{LS}$, related to the synchronisation of a connected pair of oscillators. The second is the the \textit{local desynchronisation} time $\tau_{LD}$, related to the typical desynchronisation time as the link between two synchronized oscillators is severed. The third one, finally, is the \textit{effective rewiring} time $\tau_{ER}$, describing the typical time needed for an oscillator to establish a new link after a rewiring event.

We focus on the limit in which $\tau_{LS}$ is much smaller than both $\tau_{LD}$ and $\tau_{ER}$. In this regime, oscillators couples quickly synchronize when connected by a link, starting to loose their relative synchrony when their mutual link is deleted in a rewiring event. In practice, oscillators tend to loose the information gained when linked with the characteristic timescale $\tau_{LD}$.
Two possibilities are then in order for low connectivity. Either a new link is forged by one of these two oscillators with a third node in a time shorter than $\tau_{LD}$, propagating the information it carries from its previous local synchronization to a new node, or no link at all is established before this information is completely lost. We argue that {\it global} synchronization will take place when, on average, the information gained by local synchronization events does not get lost but is rather able to propagate through the entire network. This will happen when  $\tau_{LD} \lessapprox \tau_{ER}$. On the other hand, when $\tau_{ER} \lessapprox \tau_{LD}$, no information can propagate through the network, and macroscopic synchronization cannot take place. The transition from the desynchronized to the synchronized regimes will thus take place when
\begin{equation}
\tau_{LD} \approx \tau_{ER} \,.
\label{synch_cond}
\end{equation}

Note that a similar argument, based on the characteristic timescales of information transfer, has been previously successfully applied to estimate the transition line separating disordered from collective motion in the well known Vicsek model for flocking \cite{Ginelli2008, Ginelli2016, Grygera2018}.

We now proceed to estimate the three timescales introduced above.
First consider the local synchronization scale $\tau_{LS}$, that is, the time needed by two oscillators $i$ and $j$ sharing a non-directed link to synchronize their effective frequencies. 
In the low connectivity approximation one can assume for a couple of oscillators $m^t_i=m^t_j=1$ , i.e. that they are only connected one to each other. Hence, from Eq. (\ref{maineq}), one immediately gets for their mutual phase difference $\delta \varphi=\varphi_i-\varphi_j$ the dynamics
\begin{equation}
\delta\dot{\varphi} = \delta\omega -2 \varepsilon \sin \delta\varphi \; , 
\end{equation}
where $\delta\omega=\omega_i - \omega_j$ is the difference between their natural frequencies. In the strong coupling regime we are interested into, $\varepsilon \gg \sigma $ and one readily sees that the phase difference converges exponentially fast towards the asymptotic solution $\delta\varphi =  \delta\omega /(2 \varepsilon)$ while the two oscillators effective frequencies synchronize with a time scale $\tau_{LS} \approx (2 \varepsilon)^{-1}$. In the following we first assume $\tau_{LS} \ll T$, that is, once a link is established oscillators typically synchronize before being rewired.

\begin{figure}[t!]
\centering
\includegraphics[width=0.85\linewidth]{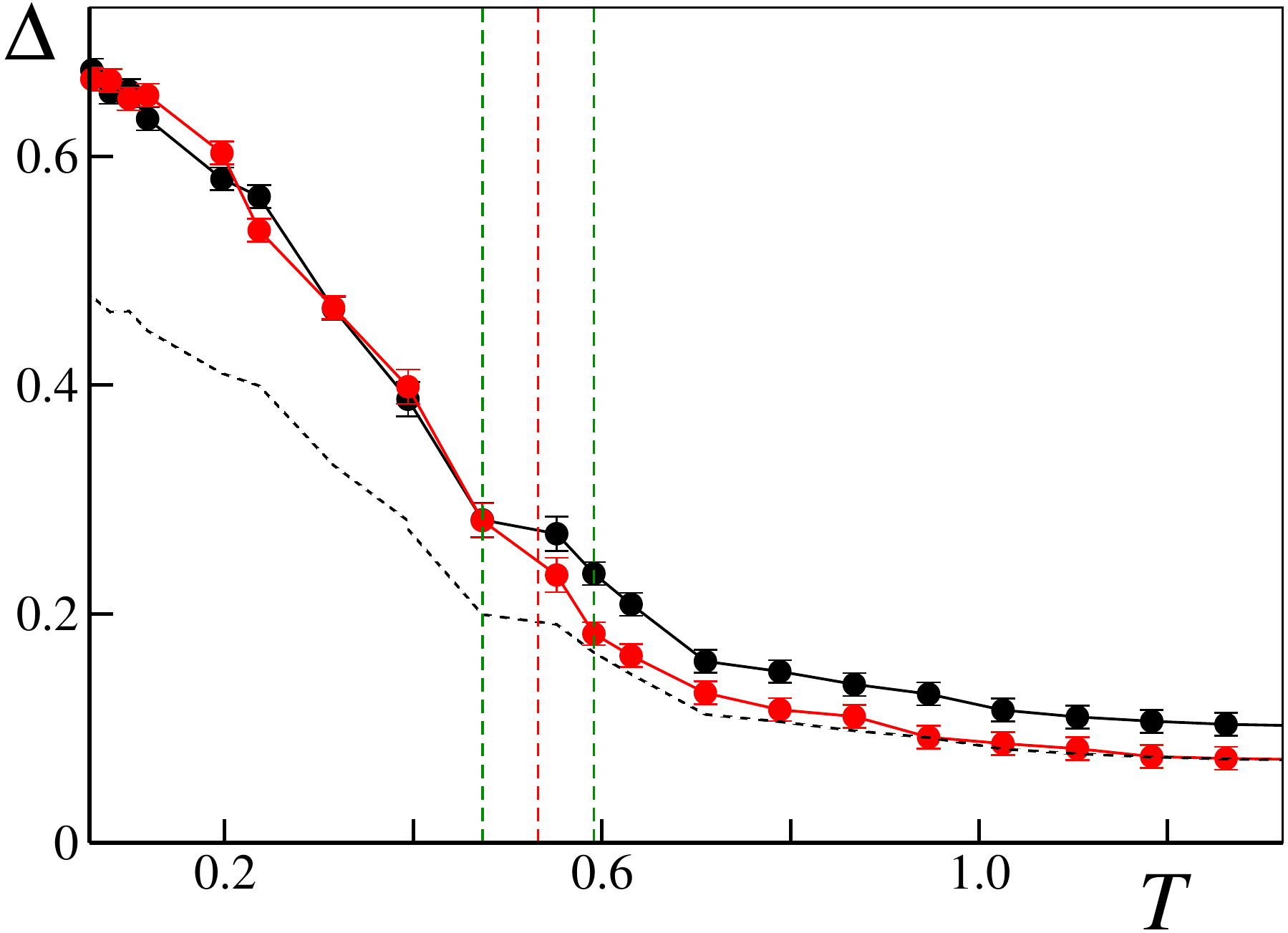}
\caption{Finite size determination of the transition point $T_c$ for $q=0.8$, $\e=8$ and $\sigma=1$. The average order parameter $\Delta(N)$ is evaluated for two different system sizes, respectively $N_1=1000$ (black dots) and $N_2=2000$ (red dots). $T_c$ is estimated as the midpoint between the largest value of $T$ such that the values of $\Delta(N_1)$ and $\Delta(N_2)$ overlap, and the smallest value of $T$ such that the scaling $\Delta(N_1)/ \Delta(N_2) \approx \sqrt{2}$ is satisfied. To facilitate the comparison, the black dashed line marks the value $\Delta(N_1)/\sqrt{2}$. Vertical dashed lines mark the estimated transition point (red) and its confidence interval (green). Error bars report the standard error for the average computed over $\Omega=20$ independent realizations. }
\label{figure4}
\end{figure}

Once the link is removed in a rewiring event, nodes can be left without any link, so that the phase of previously connected and synchronized oscillators will start to drift away one from each other due to their natural frequencies difference $\delta \omega$, loosing any information regarding their previous mutual synchronization when their phase difference approaches $\pi /2$. This allows one to define the typical local desynchronization timescale $\tau_{LD}$ such that  
\begin{equation}
\tau_D \langle \delta \omega \rangle \approx \frac{\pi}{2}
\end{equation}
with being the average natural frequency difference. For Gaussian distributed natural frequency one of course has
\begin{equation}
\langle \delta \omega \rangle =\sqrt{\int_{-\infty}^{\infty} d\omega_1d\omega_2\, P_\sigma(\omega_1)P_\sigma(\omega_2)(\omega_1-\omega_2)^2}=\sqrt{2}\sigma \; 
\end{equation}
which finally yields
\begin{equation}
\tau_{LD} \approx \frac{\pi}{2\sqrt{2}\,\sigma}
\end{equation}

Before proceeding further, one comment is in order about our estimate of the typical local desynchronization timescale. We have computed it as the time required by a typical pair of oscillators to desynchronize. This is of course different from the average of individual couples desynchronization times
$ \langle \pi /(2\delta\omega ) \rangle$,
which is dominated by oscillators couples with almost degenerate natural frequencies, $\delta \omega \approx 0$. These latters, however, characterized by a very large local desinchronization time, are far from being representative of the typical behavior of random oscillators couples.  

\begin{figure}[t!]
\centering
\includegraphics[width=0.95\linewidth]{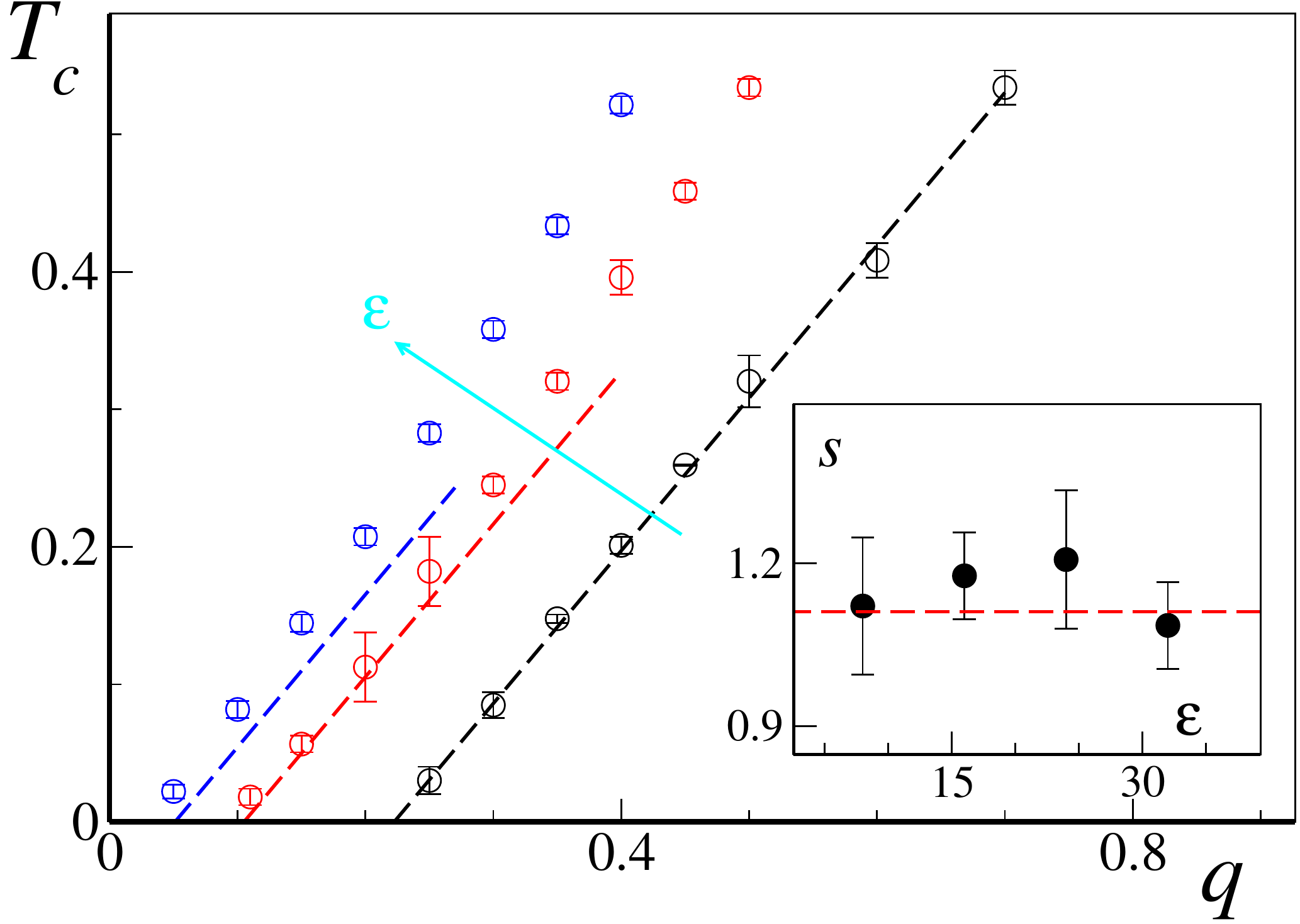}
\caption{Critical rewiring time $T_c$ as function of $q$ for $q<1$ for $\sigma=1$ and different values of the coupling constant (increasing along the cyan arrow). Respectively, from left to right: $\varepsilon=32$ (blue circles ), $\varepsilon=16$ (red circles) and $\varepsilon=8$ (black circles). Error bars give the estimated upper and lower boundaries for $T_c(q)$ as discussed in the main text. The dashed straight lines (same color coding) mark the linear prediction of Eq. \ref{eq_final} (see Section \ref{ave_net}).
(Inset): The slope $s$ of each $\e$ curve, evaluated by linear regression of the main panel data, is compared with the theoretical estimate $s= \pi / (2 \sqrt{2})$ (see Eq. (\ref{mf})). Data has been averaged over $\Omega=20$ different realizations and error bars measure one standard error.}
\label{figure3}
\end{figure}
We finally estimate the effective rewiring timescale $\tau_{ER}$. 
The probability for an oscillator to be linked to at least one other oscillator is equal to:
\begin{equation}
P_\text{link}=1-P_\text{not\ link}\,
\label{plink}
\end{equation}
where $P_\text{not\ link}$ is the probability of not having any link at all, that is
\begin{equation}
P_\text{not\ link}=\left(1-\frac{q}{N}\right)^{N-1} \xrightarrow[]{N\to\infty} e^{-q}
\end{equation}
Substituting back into Eq. (\ref{plink}) we get
\begin{equation}
P_\text{link} \approx 1-e^{-q} \approx q\;\;\;\mbox{for}\;\; q \ll 1\,,
\end{equation}
with lowest order corrections of order $q^2$ and $q/N$. We thus evaluate the effective rewiring time in the low connectivity limit as
\begin{equation}
\tau_{ER} = \frac{T}{P_\text{link}} \approx \frac{T}{q}\,.
\end{equation}

Summing up, the synchronization condition (\ref{synch_cond}) yields a linear relation between the rewiring time $T$ and the connectivity $q$, yielding the synchronization line
\begin{equation}
T_c (q)\approx \frac{\pi}{2\sqrt{2}\,\sigma}\,q
\label{mf}
\end{equation}

We now compare our predictions with numerical simulations. We determine the synchronization threshold by finite size analysis, comparing the averaged order parameter $\Delta(N)$ for system sizes $N_1=1000$ and $N_2=2000$. In the presence of macroscopic synchronization one expects $\Delta(N_1) \approx \Delta(N_2)$, while in the disordered phase we have
\begin{equation}
\frac{\Delta(N_1)}{\Delta(N_2)}=\sqrt{\frac{N_2}{N_1}}=\sqrt{2}
\end{equation}
An example of our procedure is given in Fig. \ref{figure4} for $q=0.8$, where we have estimated $T_c=0.53(6)$.

Numerical estimates of the synchronization threshold are reported in Fig.~\ref{figure3} for $\sigma=1$ and different values of the coupling constant $\varepsilon$. They confirm the linear relation between $T_c$ and $q$ in the low connectivity regime, predicting the actual slope $s=\pi/(2\sqrt{2}\, \sigma )$ within numerical accuracy (see inset). 
However, it is clear that for finite values of the coupling, it is always possible to find sufficiently small values of $q$ such that synchronization cannot be achieved, no matter how small is $T$. Said differently, the critical line $T_c$ has a non-zero intercept $q_0 (\e)$ with the $T=0$ axis. Interestingly, the value of $q_0(\e)$ of the intercept decreases towards zero as $\varepsilon$ increases, suggesting that Eq. (\ref{mf}) can be fully recovered as $\e \to \infty$. 

This is equivalent to the strong coupling limit under which we have derived Eq. (\ref{mf}): By taking the limit $\e \to \infty$ first, in fact, we assure that the condition
$\tau_{LS} = (2 \e)^{-1} \ll T$ is verified for any non-zero rewiring time $T$. On the other hand, numerical simulations with a finite coupling constant $\e$ show that one can always find a sufficiently low connectivity $q$ such that $T_c(q) \lessapprox \tau_{LS} = (2 \e)^{-1} $ and our approximation breaks down.

In the next section, we will attempt to better understand this regime and the behavior of the  intercept $q_0 (\e)$ by means of averaging considerations.

%---------------------------------------------------------------------------------

\subsection{Average network for very fast rewiring}
\label{ave_net}
We now consider the limit of extremely fast rewiring, where $T$ and $\tau_{ER} \approx T/q$ are much smaller than the local synchronization and desynchronization times. In this regime, one expects the instantaneous order parameter to be approximately constant over a timescales $\tau_{av} \lessapprox \mbox{min}(\tau_{LS},\tau_{LD})$, so that
\begin{equation}
R(t) \approx \frac{1}{\tau_{av}} \int_t^{t+\tau_{av}} R(t+t')\, dt' \,.
\end{equation}
Following the argument of Ref. \cite{So08}, we may invoke a well known result from Ott and Antonsen \cite{Ott-Antonsen} to argue that the low dimensional dynamics of the Kuramoto order parameter is essentially controlled by the time-averaged interaction matrix
\begin{equation}
\left\langle \frac{\mathcal{A}_{ij}^t(T)}{m_i^t} \right\rangle \equiv \frac{1}{\tau_{av}} \int_0^{\tau_{av}}\frac{\mathcal{A}_{ij}^t(T)}{m_i^t} \,dt \,.
\label{ave}
\end{equation}
This result, stating that for $T \to 0$ the dynamics of Eq. (\ref{maineq}) is the same as the one of the time-average network with stationary connectivity, can be essentially seen as a form of the averaging theorem \cite{AveragingT}. While the latter typically involves periodic systems, a recent extension to non-periodic systems has been discussed, for instance, in Ref. \cite{Duccio}.

In the limit $T \to 0$ the average in Eq. (\ref{ave}) is computed over arbitrarly many rewiring events and we have
\begin{equation}
\frac{A_{ij}}{N} \equiv \lim_{T \to 0} \int_0^{\tau_{av}}\frac{\mathcal{A}_{ij}^t(T)}{m_i^t} \,dt = \sum_k^{N-1} \frac{a_{ij} (k)}{k}\,,
\label{avA}
\end{equation}
where 
\begin{equation}
a_{ij} (k) = p^k (1 - p)^{N-1-k} \frac{(N-2)!}{(k-1)!(N-k-1)!}
\label{avB}
\end{equation}
is the probability that node $i$ has an active link with node $j$ and exactly $k-1$ other links. Note that the binomial factor
\begin{equation}
\binom{N-2}{k-1}  \equiv \frac{(N-2)!}{(k-1)!(N-k-1)!}
\end{equation}
accounts for all the different configurations in which the $k-1$ active link can be chosen out of $N-2$ potential ones after the one between  $i$ and $j$ has been activated.

By recalling that $p=q/N$, using Eqs. (\ref{avA})-(\ref{avB}) one can find 
\begin{eqnarray}
\frac{A_{ij}}{N} &=& \sum_k^{N-1} \frac{q^k}{k\,(k-1)!} \frac{1}{N^k} \frac{(N-2)!}{(N-k-1)!} \left(1 - \frac{q}{N}\right)^{N-k-1}\nonumber\\
&=&\left(1 - \frac{q}{N}\right)^{N}\frac{1}{N}\sum_k^{N-1} \frac{q^k}{k!} \,g(N,k) \,,
\end{eqnarray}
where
\begin{equation}
g(N,k)=\frac{1}{N^{k-1}}\frac{(N-2)!}{(N-k-1)!} \left(1 - \frac{q}{N}\right)^{-(k+1)}\,.
\end{equation}
In the limit $N \gg 1$ we have 
\begin{equation}
g(N,k) = 1 + O\left(\frac{1}{N}\right)
\end{equation}
and therefore, to lowest order in $1/N$,
\begin{equation}
\frac{A_{ij}}{N}\approx \frac{e^{-q} }{N}\sum_k^{\infty} \frac{q^k}{k!} \approx \frac{e^{-q}}{N} \left( e^q -1\right) = \frac{1 - e^{-q}}{N}
\end{equation}
so that the average network is characterized by a globally connected topology.

Therefore, under our conjecture, in large networks the fast rewiring ($T\to 0$) dynamics (\ref{maineq}) can be replaced by the averaged one
\begin{equation}
\frac{\partial \p_i}{\partial t} = \om_i + \frac{\varepsilon\left(1 - e^{-q}\right)}{N} \sum_{j=1}^N \sin(\p_j - \p_i) \; ,
\label{aveq}
\end{equation}
that is, a globally coupled Kuramoto model with coupling constant
\begin{equation}
J=\e\left(1 - e^{-q}\right) \,,
\end{equation}
which exhibits macroscopic synchronization for $J>J_c$, with the critical point $J_c$ depending on the natural frequency distribution. For $T\to 0$, therefore, macroscopic synchronization can only be achieved provided $\e>J_c$ and for connectivities $q>q_0$ with
\begin{equation}
q_0 = \ln \left(\frac{\e}{\e - J_c}\right) \,.
\label{q0}
\end{equation}

We conclude that, according to Eq. (\ref{q0}), in the strong coupling limit $\e/\sigma \to \infty$ synchronization can be achieved for arbitrarily small connectivity $q$.

\begin{figure}[t!]
\centering
\includegraphics[width=0.95\linewidth]{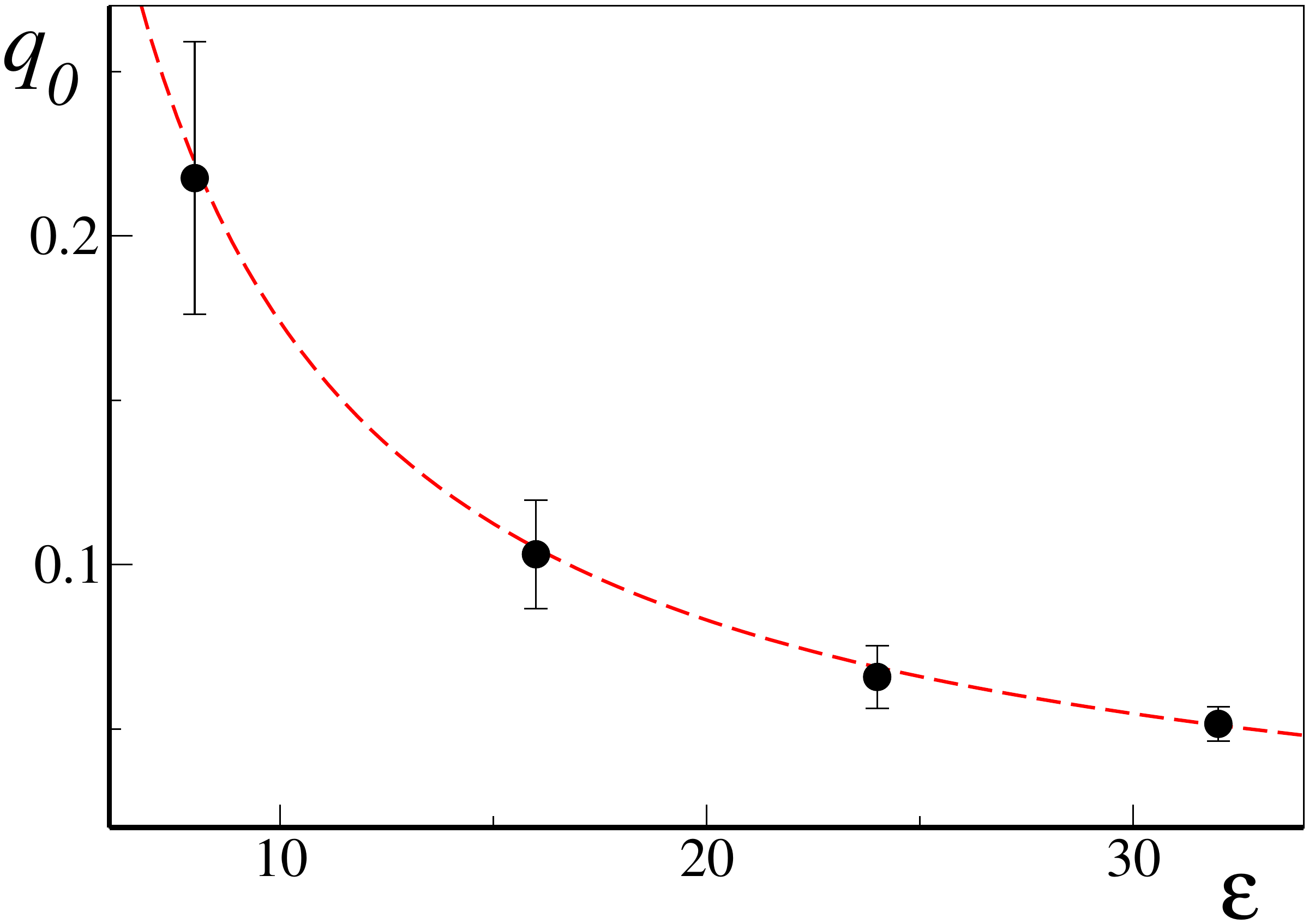}
\caption{Numerically estimated transition line intercepts $q_0$ as a function of the coupling constant $\e$ (black dots) are compared with the analytical prediction given by Eq. (\ref{q0}) (dashed red line). Data as in Fig.~\ref{figure3}. Error bars represents error in the linear extrapolation process (see main text).}
\label{figure5}
\end{figure}

Furthermore, we can interpret $q_0$ as the intercept of the transition line $T_c(q_0)$ with the $T=0$ axis. In particular, for a Gaussian distribution of natural frequencies with unit standard deviation we have $J_c = \sqrt{8/\pi}$ \cite{Strogatz2000}, which allows us to to compare Eq. (\ref{q0}) with the intercept values obtained by extrapolating the best linear fit for the transition lines of Fig.~\ref{figure3}. Direct comparison (see Fig.~\ref{figure5}) shows excellent agreement in the coupling range $\e \in [8,32]$ we have probed. 
We can now correct Eq. (\ref{mf}) by adding a constant term such that $T_c(q_0) = 0$, thus obtaining
\begin{equation}
T_c (q)\approx \frac{\pi}{2\sqrt{2}\,\sigma}\,(q - q_0) 
\label{eq_final}
\end{equation}
with
\begin{equation}
q_0 = \ln \left(\frac{\e}{\e - \sqrt{8/\pi}}\right) \,,
\label{q00}
\end{equation}
for Gaussian distributed natural frequencies.
This is exactly the linear formula we plotted in Fig.~\ref{figure3} for unit variance  ($\sigma=1$), showing good comparison with the numerical transition values $T_c(q,\e)$ in the small $T$ regime. As $q$ (and thus $T_c$) grows larger, however, deviations from the linear behavior are clearly visible. Indeed, as $q$ is increased, the critical line $T_c(q)$ bends upwards to meet the vertical asymptote at $\bar{q}$. In this regime, contributions from nodes with more than one link at the time becomes relevant, and the simple arguments leading to the linear relation (\ref{eq_final}) are expected to break down.

%------------------------------------------------------------------------------------------------------------------------------------------------

\section{Behavior at finite connectivity}
\label{sec4}
For completeness, in this section we briefly discuss the synchronization transition at finite connectivity $q$.

\subsection{Behaviour for large connectivity}
\label{largeC}

We have already seen that, in order to synchronize for arbitrarily large rewiring times, the connectivity $q$ should be larger than a threshold $\bar{q} > 1$, so that the typical emergences of connected components of macroscopic size, taking place for $q>1$, is not sufficient for the onset of synchronization. In particular, we have seen that for $\e/\sigma=8$ we have  $\bar{q} = 1.66(6)$. We now show numerical evidence that in the large coupling limit, $\e/\sigma \to \infty$, we have $\bar{q} = 1^+$, that is the onset of synchronization do coincide with the emergence of giant connected components in the graph topology.

\begin{figure}[t!]
\includegraphics[width=0.95\linewidth]{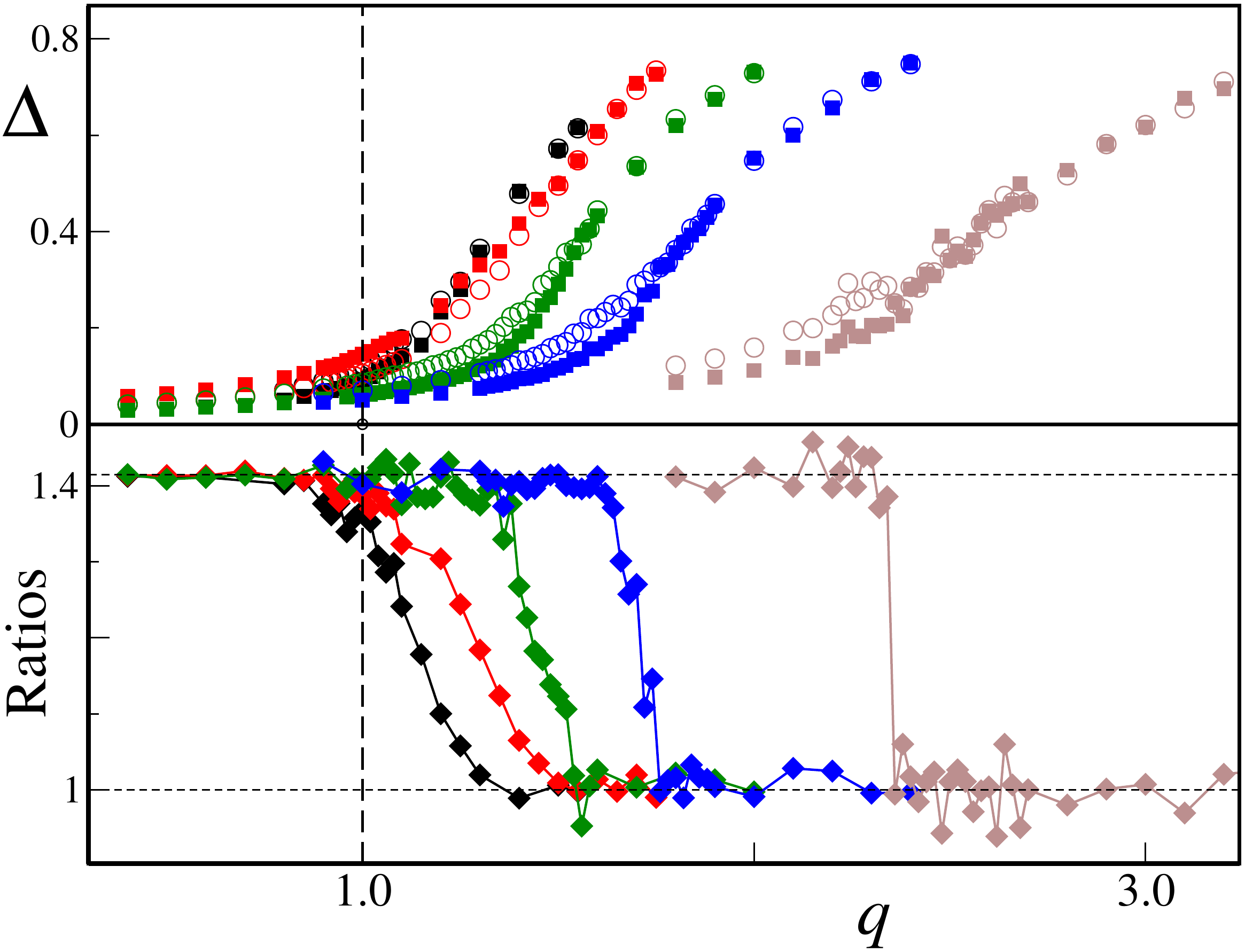}
\centering
\caption{Stationary values $\Delta$ of the system as function of $q$ for $T=20\pi/\sigma$. 
Top: Behavior of $\Delta$ as a function of $q$ for two different sizes ($N_1=10^3$, empty circles, and $N_2=2\times 10^3$, full squares) and different values of $\e/\sigma$. 
From the left to the right: $\sigma^2=1/8$ (black), $\sigma^2=1/4$ (red), $\sigma^2=1/2$ (green), $\sigma^2=1$ (blue) and $\sigma^2=2$ (brown).
Bottom: Ratios $\Delta(N_1)/\Delta(N_2)$ as a function of $q$. The two horizontal dotted lines mark the ratios $\sqrt{2}$ (disordered phase) and 1 (synchronized phase).
The colours coding for the variance is the same as in the top panel. The vertical dashed line $q=1$ marks the emergence of giant connected components.
Data has been averaged over $\Omega=10$ different realizations.}
\label{FigSpread}
\end{figure}

Next we show that the phase transition thresholds depend on the distribution of frequencies $\omega$ when the coupling $\varepsilon$ is fixed, which demonstrates that the critical point is not dependent on the topology (and the percolation threshold for the giant connected component), but only on the dynamics. 

In the following, we analyze numerically the synchronization transition at in the region $q \approx 1$ through the finite size analysis of the averaged parameter $\Delta$ at large rewiring times $T$ as the ratio  $\e/\sigma$ is progressively increased. Fixing $\e$, we increase $\sigma$ between $1/\sqrt{8}$ and $\sqrt{2}$. In order to evaluate $\Delta$ at large enough rewiring times, in agreement with the scaling relation (\ref{scaling}) we fix $T$ such that the dimensionless parameter $T \sigma = 20 \pi$ \cite{NOTE2}, and compare the order parameter at two different system sizes $N_1 < N_2$. As already remarked in Sec. \ref{scales}, we can distinguish the synchronized from the disordered phase by the ratio $\Delta(N_1)/\Delta(N_2)$. This is, for instance, how in Sec.~\ref{sims} we have estimated $\bar{q}=1.66(6)$ for $\e/\sigma=8$ from the data of Fig.~\ref{FigSpread} (blue symbols). 

More in general, numerical simulations, reported in Fig.~\ref{FigSpread}, clearly indicates that, as $\sigma$ is lowered and the strong coupling regime is approached, the synchronization threshold approaches the onset of giant connected components, i.e. $\bar{q} \to 1$. 
These results indicate that for finite couplings, in the regime $1<q<\bar{q}$, giant connected components may be unable to synchronize when large enough rewiring times $T$ are considered. 
This effect is indeed due to the interaction between the giant component topology and the quenched disorder. For $q  \gtrapprox 1$ the giant component should be characterized by a large number of bridges (i.e. links whose deletion would split the giant component in two disconnected parts). When these bridges insist on nodes characterized by extreme natural frequencies (i.e. lying in the tail of the distribution $P(\omega)$) which do escape partial synchronization, they act as effective obstacles to information spreading, splitting the topologically connected giant component into different synchronized subcomponents which are, however, not  mutually synchronized. This mechanism, which clearly prevent macroscopic synchronization to emerge in the slow switching regime, is however going to become less and less important as the connectivity $q$ is increased and the number of bridges in the giant connected component is reduced, allowing for a more efficient information flow, eventually leading to global synchronization as $q > \bar{q}$. 

Fast switching, on the other hand, allows information to travel through the network by rearranging the giant cluster quickly enough, preventing instantaneous bridges from acting as effective roadblocks. Therefore, for $1<q<\bar{q}$, a transition to macroscopic synchronization is eventually observed as the rewiring time is decreased. A precise analytical estimate of $T_c(q)$ in this regime, however, is beyond the scope of this work.

\subsection{Critical behavior}

\begin{figure}[t!]
\includegraphics[width=0.95\linewidth]{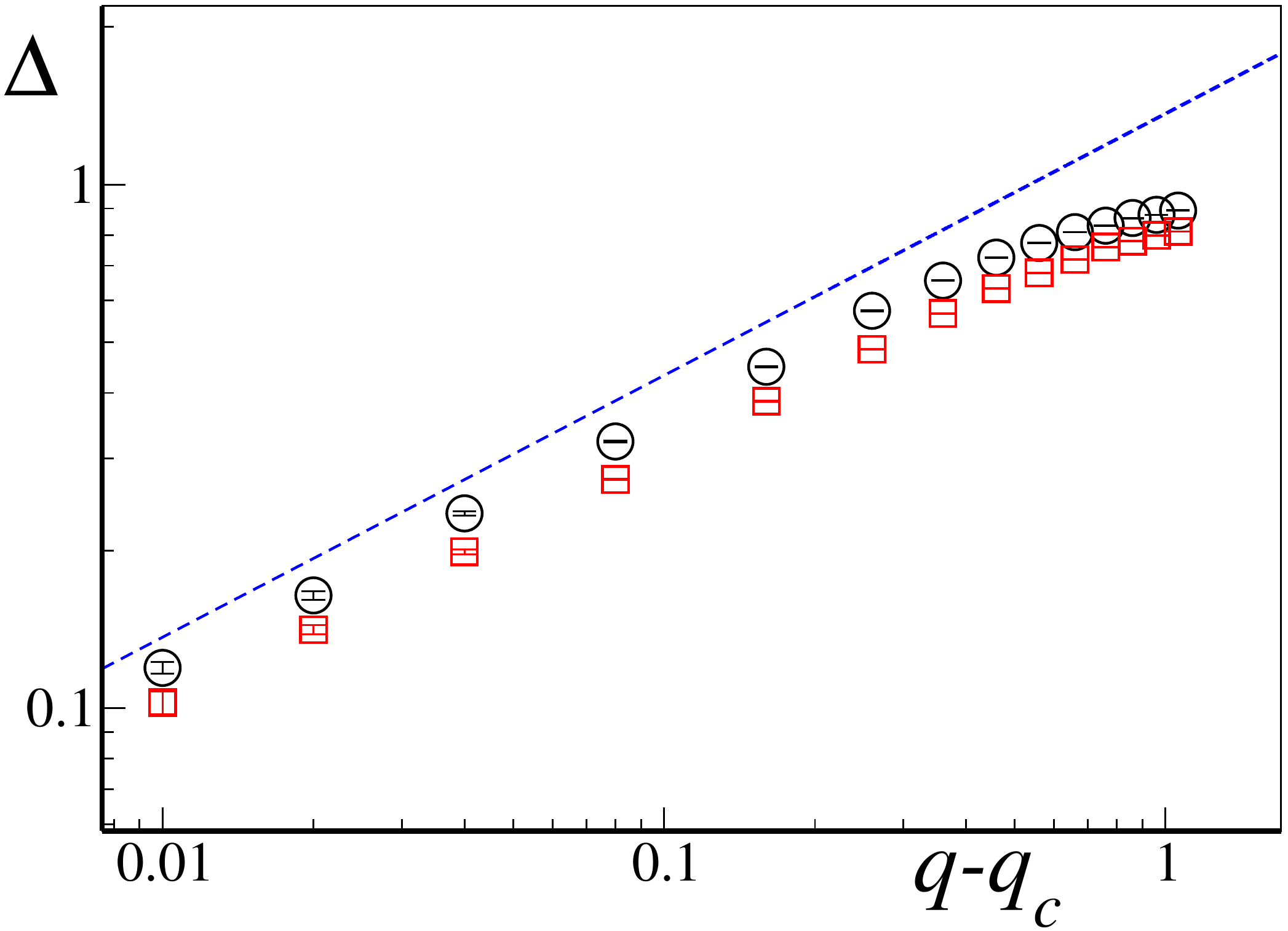}
\centering
\caption{Log-log plot for the critical behavior of $\Delta\sim (q-q_c)^\beta$ in a network with $N=10^4$, $\e=8$ and $\sigma=1$ for faster ($T=0.31$, black circles) and slower ($T=9.42$, red squares) switching times. The blue dashed curve marks the Kuramoto exponent $\beta=1/2$. Data has been averaged over $\Omega=20$ independent realization of the natural frequencies.}
\label{critp}
\end{figure}

We have finally verified that, as expected, the phase transition to synchronization belongs to the usual Kuramoto model class. 
In Fig.~\ref{critp} we report numerical simulations for both (relatively) slow and fast rewiring times $T$, showing that the average order parameter follows the usual  
Kuramoto model scaling, $\Delta \sim \sqrt{q -q_c(T)}$ for $q > q_c$ \cite{Strogatz2000}, with $q_c(T)$ being the ($T$ dependent) critical connectivity parameter.
Numerical results suggest this to be true for any finite $q_c$, as expected given that synchronization seems to be essentially guided by the properties of the globally connected time-averaged connectivity matrix.

%---------------------------------------------------------------------------------
\section{Conclusions}
\label{sec5}

We discussed a time-varying network of heterogeneous Kuramoto phase oscillators characterized by links being randomly switched on and off with a Poissonian probability distribution. The network dynamics exhibits three well defined time scales associated respectively to local synchronization, local desynchronization and effective rewiring, whose separation is controlled by model parameters. 
Numerical simulations and analytical arguments show that this system is able to achieve  statistically stable macroscopic synchronization even for arbitrarily small net connectivity (i.e., for a sparse and infrequent coupling among the oscillators), provided sufficiently fast switching and strong couplings are considered. 

In the formal fast switching limit, $T \to 0$, we have argued that the synchronization dynamics is fully captured by the time-averaged connectivity matrix, suggesting that results from the averaging theorem can be applied to our Kuramoto setup. At finite $T$, on the other hand, our analytical arguments, based on the comparison between the different timescales at play, are indeed able to predict with a satisfactory precision the synchronization transition line in the small connectivity regime. 
This switching-induced synchronization maintains the same qualitative characteristics of its static counterpart, such as the Kuramoto order parameter scaling $\Delta \sim \sqrt{q-q_c}$
at $q \gtrapprox q_c$ \cite{acebron}. 

For larger connectivity values, beyond the onset of giant connected components ($q>1$), we have finally shown that the interaction between instantaneous connectivity topology and quenched disorder  may prevent the onset of synchronization for sufficiently slow rewiring times and large but finite couplings. 

Our findings are primarily intended as a theoretical contribution to the field of synchronization in time-varying complex networks, and in particular non-equilibrium synchronization models with alternative mechanisms giving rise to synchronization. However, we can still envisage several lines of potential applications for our results. For instance, one can think of engineered systems of heterogeneous (and possibly mobile) units \cite{mobileoscillators} -- a simple paradigma for the "Internet of Things" \cite{iot} -- where maintaining constant connectivity could be costly, yet the system is still required to exhibit synchronization or other collective properties. Our model could help develop alternatives to constant interactions, able to generate the same collective dynamics albeit a sparse and seldom connectivity. Also, as mentioned earlier, this setup could be seen as a crude model for social interactions. As such, our model could be used to qualitatively model the emergence of consensus in a community where different individuals are only interacting with a few of their contacts at any time.

Finally, these results open several avenues of future work. Rather than rewiring links at random, one can consider a network where links are rewired preferentially to nodes with a similar instantaneous dynamical state, thus favoring interactions with ``like-minded'' individuals. This set-up could be used, for instance, to investigate qualitatively the ``echo-chambers'' phenomenon in social media, which as been recently suggested to be a possible source of an increased polarization in political opinions. To this regard, one can also wish consider different distribution of quenched frequencies, such as uniformly distributed ones, which do not favour middle natural frequencies (i.e. opinions) as the Gaussian one.

%---------------------------------------------------------------------------------

\vspace{0.3 cm}
\section*{Acknowledgments}
We wish to thank D. Fanelli and M. Lucas for fruitful discussions. This work has been supported by H2020-MSCAITN-2015 Project COSMOS No. 642563. ZL also acknowledges support from "Slovenian research agency" via P1-0383 and J5-8236". FR acknowledges support from H2020 MSCA grant agreement No. 702981.

%\bibliography{your-bib-file}

\begin{thebibliography}{99}

\bibitem{VespignaniBook} A. Barrat, M. Barth\'elemy, A. Vespignani, {\it Dynamical Processes on Complex Networks} (Cambridge University Press, Cambridge, 2008).
\bibitem{newman} Newman M {\it Networks: an Introductio} (Oxford University Press, Oxford 2010).
\bibitem{costa} L. da Fontoura Costa {\it et al.}, Adv. in Phys. {\bf 60}, 329 (2011).
\bibitem{barabasi} A. L. Barabasi, {\it Network Science} (Cambridge University Press, 2016).

\bibitem{mason} M. Porter and J. Gleeson, {\it Dynamical Systems on Networks} (Springer Verlag, Berlin 2016).
\bibitem{arkadybook} A. Pikovsky, M. Rosenblum, and J. Kurths. {\it Synchronization: A universal  concept in nonlinear sciences} (Cambridge University Press, Cambridge, 2001).
\bibitem{Arenas2008} A. Arenas, A. D\'iaz-Guilera, J.Kurths, Y. Moreno, C. Zhou, Phys. Rep. {\bf 469} 93 (2008).
\bibitem{arkadyandme}  Z. Levnajić, A. Pikovsky, Physical Review E {\bf 82}, 056202 (2010).
\bibitem{arkadymisha} A. Pikovsky, M. Rosenblum, Chaos {\bf 25}, 097616	(2015).

\bibitem{Rodrigues2016} 
F. A. Rodrigues, T. K. DM. Peron, P. Ji, J. Kurths, Phys Rep {\bf 610} 1 (2016).

\bibitem{acebron} J. A. Acebrón, Rev. Mod. Phys. {\bf 77}, 137 (2005).

\bibitem{Ermentrout91}
B. Ermentrout, J. Math. Biol. {\bf 29} 571 (1991).

\bibitem{Ballerini08}
M. Ballerini, et al., Proc. Natl. Acad. Sci. USA {\bf 105}, 1232 (2008).

\bibitem{Markram97}
H. Markram, J. L\"ubke, M. Frotscher, B. Sakmann, Science {\bf 275} 213 (1997).

\bibitem{Maistrenko07}
Y. L. Maistrenko {\it et al.}, Phys. Rev. E {\bf 75} 066207 (2007).

\bibitem{Pini2011}
G. Pini, A. Brutschy, M.Frison, A.Roli, M. Dorigo, M.Birattari, Swarm Intelligence {\bf 5} 283 (2011).

\bibitem{Sekara16}
V. Sekara, A. Stopczynski, and S. Lehmann, Proc. Natl. Acad. Sci USA  {\bf 113}  9977 (2016). 

\bibitem{Hua2009}
H. Hua, S. Myers, V. Colizza, A. Vespignani, Proc. Natl. Acad. Sci USA {\bf 106} 1318 (2009).

\bibitem{Hasler04}
I.V. Belykh, V.N. Belykh, M. Hasler, Phys D  {\bf 195} 188 (2004).

\bibitem{Stilwell06}
D. J. Stilwell, E. M. Bollt, and D. G. Roberson SIAM J. Appl. Dyn. Syst., {\bf 5} 140 (2006).

\bibitem{Amritkar06} 
R. E. Amritkar, and Chin-Kun Hu, Chaos {\bf 16}, 015117 (2006).

 \bibitem{Li08} Z. Li, L. Jiao, J. Lee, Physica A {\bf 387} 1369 (2008).
 %Robust adaptive global synchronization of complex dynamical networks by adjusting time-varying coupling strenght, 

\bibitem{Lucas18}
M. Lucas, D. Fanelli, T. Carletti, and J. Petit, Europhys. Lett. {\bf 121} 50008 (2018).

\bibitem{Strogatz2000}
S.H. Strogatz, Phys D. {\bf 143} 1 (2000).

\bibitem{AveragingT}
F. Verhulst, {\it Nonlinear Differential Equations and Dynamical
Systems} (Springer Science \& Business Media, 1990).

\bibitem{So08}
P. So, B. C. Cotton, and E. Barreto, Chaos {\bf 18} 037114 (2008).

\bibitem{kuramoto1}Kuramoto,  Y.,  1975, {\it  International  Symposium  on  Mathematical Problems in Theoretical Physics, Lecture Notes in Physics No. 30} Springer, New York, p. 420. (2005)
 
\bibitem{kuramoto2}Kuramoto, Y., {\it Chemical Oscillations, Waves and Turbulence}, Springer, New York (1984)

\bibitem{NOTE1} Note that the instantaneous dynamics itself is not fully symmetric, as two nodes $i$ and $j$ sharing the same link can still have different instantaneous degrees.

\bibitem{giantcomponent} S.N. Dorogovtsev and A.V. Goltsev, Review of Modern Physics, Vol.{\bf 80} 1275 (2008).
 
\bibitem{Ginelli2008} H. Chat\'e, F. Ginelli, G. Gr\'egoire and F. Raynaud, Phys Rev E, {\bf 77}, 046113 (2008).

\bibitem{Ginelli2016} F. Ginelli, Eur. Phys. J. Spec. Top. {\bf 225} 2099  (2016)

\bibitem{Grygera2018} M Puzzo, A. De Virgiliis, TS Grigera, arXiv:1810.02141 (2018).

\bibitem{Ott-Antonsen} E. Ott, and T. Antonsen, Chaos {\bf 18}, 037113 (2008).

\bibitem{Duccio} J. Petit, B. Lauwens, D. Fanelli and T. Carletti, Phys. Rev. Lett. {\bf 119}, 148301 (2017)

\bibitem{NOTE2} We have verified that larger choices for $T \sigma$ do not change qualitatively the overall picture. 

\bibitem{mobileoscillators} N. Fujiwara, J. Kurths, A. Diaz-Guilera, Phys. Rev. E  {\bf 83}, 025101(R) (2011); N. Fujiwara, J. Kurths, A. Diaz-Guilera, Chaos {\bf 26}, 094824 (2016).

\bibitem{iot}J. Gubbi, R. Buyya, S. Marusic and M. Palaniswami, Future Gener. Comput. Syst. {\bf 29}, 1645--1660 (2013)


\end{thebibliography}

\end{document}